\documentclass[11pt,a4paper,english,showpacs]{revtex4}
\usepackage{amssymb}
\usepackage{amsmath}
\usepackage{graphicx}
\usepackage{babel}
\usepackage{hyperref}

\begin{document}

\title{Dynamical generation of mass in the $D=(2+1)$ Wess-Zumino model}

\author{A.~C.~Lehum}
\email[]{lehum@fma.if.usp.br}
\affiliation{Instituto de F\'{\i}sica, Universidade de S\~{a}o Paulo\\
 Caixa Postal 66318, 05315-970, S\~{a}o Paulo - SP, Brazil}

\date{\today}

\begin{abstract}

In this work we study the dynamical generation of mass in the massless ${\cal N}=1$ Wess-Zumino model in a three dimensional spacetime. Using the tadpole method to compute the effective potential, we observe that supersymmetry is dynamicaly broken together with the discrete symmetry $A(x)\rightarrow -A(x)$. We show that this model, differently from non-supersymmetric scalar models, exhibits a consistent perturbative dynamical generation of mass after two loop corrections to the effective potential.

\end{abstract}

\maketitle

 Supersymmetry (SUSY) is one of the most beautiful theoretical accomplishments of the present time. However, to construct realistic models of particle physics, SUSY should be spontaneously broken~\cite{Witten:1982df} at some energy scale, since we do not observe the superpartners of the ordinary particles (e.g. seletrons, squarks, etc...). Due to the non-renormalization theorem, however, spontaneous breaking of SUSY is ruled out for a wide class of models~\cite{Witten:1982df}. Theoretically, it is essential to understand under which conditions SUSY can be broken, while it is the experimental data that will eventually tell us how SUSY is broken.

Toy models constructed in three dimensional spacetime have been intensely explored in the literature as a good theoretical laboratory. As indicated by Witten~\cite{Witten:1981nf}, differently from models defined in four dimensional spacetime, dynamical SUSY breaking can occur in $D=2+1$ spacetime, and this possibility was studied for the Wess-Zumino model and massless electrodynamics at one loop order, showing that no breaking appears~\cite{Burgess:1983nu}, i.e., up to one loop order neither SUSY nor gauge symmetry are dynamically broken. However, an investigation of higher loop order is still necessary to verify whether dynamical SUSY breaking really happens in such models. 

On the other hand, dynamical generation of mass in a non-supersymmetric purely scalar model is inconsistent with perturbation theory~\cite{Coleman:1973jx,Dias:2003pw,Tan:1996kz,Tan:1997ew}. In this work, we will show that, in a massless three-dimensional Wess-Zumino model, there is a consistent perturbative dynamical generation of mass induced by the discrete symmetry ($A(x)\rightarrow -A(x)$) and supersymmetry breaking.

Let us to consider the three-dimensional massless superscalar ${\cal N}=1$ Wess-Zumino model defined by
\begin{eqnarray}\label{eq1}
S=\int{d^5z}\left[\frac{1}{2}\Phi(x,\theta) D^2\Phi(x,\theta)+g\Phi(x,\theta)^4+\mathcal{L}_{CT}\right]~,
\end{eqnarray}

\noindent
where $\mathcal{L}_{CT}$ is the counter-term Lagrangian. We are using the notations and conventions as in~\cite{Gates:1983nr}.

The real superfield $\Phi$ expanded in a Taylor series in $\theta$ is given by
\begin{eqnarray}\label{eq2}
\Phi(x,\theta) = A(x)+\theta^{\alpha}\psi_{\alpha}(x)-\theta^2~F(x)~,
\end{eqnarray}

\noindent
thus, after integration in $\theta$, Eq.(\ref{eq1}) can be cast as
\begin{eqnarray}\label{eq3}
S&=&\int{d^3x}\left[\frac{1}{2}A\Box A +\frac{1}{2}\psi^{\alpha}i{\partial_{\alpha}}^{\beta}\psi_{\beta}+\frac{1}{2}F^2 
+ 4g FA^3 + 6g A^2\psi^{\alpha}\psi_{\alpha}+\mathcal{L}_{CT} 
\right]~.
\end{eqnarray}

\noindent
By eliminating the auxiliary field $F$ using its classical equation of motion, Eq.(\ref{eq3}) can be cast as
\begin{eqnarray}\label{eq3a}
S&=&\int{d^3x}\left[\frac{1}{2}A\Box A +\frac{1}{2}\psi^{\alpha}i{\partial_{\alpha}}^{\beta}\psi_{\beta}-8g^2A^6
+ 6g A^2\psi^{\alpha}\psi_{\alpha}+\mathcal{L}_{CT} 
\right]~.
\end{eqnarray}

\noindent
We can observe that the action above is invariant under the discrete symmetry transformation $A(x)\rightarrow -A(x)$.

In three dimensional theories, differently from four-dimensional models where the zero energy of the vacuum cannot be moved to a non-vanishing value by radiative corrections, it is possible in principle to break supersymmetry dynamically. To understand this issue, let us contrast the $4D$ and $3D$ effective potentials and their conditions of minimum. 

In $4D$ models, the effective potential assume the form~\cite{Grisaru:1979wc}
\begin{eqnarray}\label{eq3aa}
V_{eff(4D)}&=&-\bar{F}F-F\frac{\partial P(A)}{\partial A} +\bar{F}\frac{\partial\bar{P}(\bar{A})}{\partial\bar{A}}+\bar{F}F~G(A,\bar{A},F,\bar{F})~,
\end{eqnarray}

\noindent
where $G(A,\bar{A},F,\bar{F})$ is a function obtained by considering the radiative corrections, $P(A)$ and  $\bar{P}(\bar{A})$ are the superspace potentials evaluated in $A$ and $\bar{A}$, respectively. The origin of the form of $V_{eff(4D)}$ come from the fact that, in  perturbation theory, the effective action is obtained with a $d^4\theta=d^2\theta~d^2\bar{\theta}$ integral. Considering a classical constant superfield of the form $\Phi=A-\theta^2 F$, we conclude that the quantum corrections always come with a multiplicative $F$ and $\bar{F}$ factors.

The conditions that minimize the $4D$ effective potential are given by
\begin{eqnarray}\label{eq3bb}
\frac{\partial V_{eff(4D)}}{\partial A}&=& -F\frac{\partial^2 P(A)}{\partial A^2}
+\bar{F}F\frac{\partial G}{\partial A}=0~,\nonumber\\
\frac{\partial V_{eff(4D)}}{\partial \bar{F}}&=& -F-\frac{\partial\bar{P}(\bar{A})}{\partial\bar{A}}+FG
+\bar{F}F\frac{\partial G}{\partial \bar{F}}=0~.
\end{eqnarray}

\noindent
We can see that, if at the classical level there is some value of $A$ that satisfy $\partial P(A)/\partial A~=~0$, then $F=0$ satisfy the condition of extremum for the effective potential. It is obvious from Eq.(\ref{eq3bb}) that radiative corrections does not change this result, because in $4D$ models radiative quantum corrections are bilinear in the auxiliary fields~\cite{Grisaru:1979wc}. Thus, the minimum of the classical potential is still the minimum of the quantum effective potential, so that in $4D$ spacetime supersymmetry cannot be broken by quantum corrections. 

In three dimensional spacetime this situation can be different, nevertheless. The $3D$ effective potential assume the form
\begin{eqnarray}\label{eq3cc}
V_{eff(3D)}&=&-\frac{1}{2}F^2-F\frac{\partial P(A)}{\partial A}+F~G(A,F)~.
\end{eqnarray}

\noindent
Therefore, the radiative quantum corrections to the effective potential exhibit a single $F$ factor, which will contribute in a very different way to the conditions that minimize the effective potential. The conditions that minimize the $3D$ effective potential are given by:
\begin{eqnarray}\label{eq3dd}
\frac{\partial V_{eff(3D)}}{\partial A}&=& -F\frac{\partial^2 P(A)}{\partial A^2}
+F\frac{\partial G(A,F)}{\partial A}=0~,\nonumber\\
\frac{\partial V_{eff(3D)}}{\partial F}&=& -F-\frac{\partial P(A)}{\partial A}+G(A,F)+F\frac{\partial G(A,F)}{\partial F}=0~.
\end{eqnarray}

\noindent
In $3D$ spacetime, the superspace integrations are $d^2\theta$ and the superfields possess the form $\Phi=A-\theta^2F$. When taking derivatives with respect to $F$, the factor in
front of the term that represent the radiative corrections may disappear, and we can find that the classical minimum ($F=0$) no longer needs be a minimum of the quantum effective potential. In conclusion, $3D$ supersymmetric models may exhibit dynamical supersymmetry breaking, and our aim in this work is to show that this effectively happens. 

To study the possibility of dynamical SUSY breaking, let us dislocate the components $A$ by $v_{a}$ and $F$ by $v_{f}$, which will be interpreted as the constant vacuum expectation values of $A$ and $F$ fields at the minimum of the effective potential, respectively. In this way, Eq.(\ref{eq3}) can be written as
\begin{eqnarray}\label{seq4}
S   &=&\int{d^3x}\Big[\frac{1}{2}A\Box A +12gv_{f}v_{a}~A^2     
    +\frac{1}{2}\psi^{\alpha}i{\partial_{\alpha}}^{\beta}\psi_{\beta}
    +6gv_{a}^2\psi^{\alpha}\psi_{\alpha}
    +\frac{1}{2}F^2\nonumber\\
    &&+12g v_{a}^2 FA +12gv_{a} FA^2 + 4gFA^3 +4gv_{f} A^3+ 6g A^2\psi^{\alpha}\psi_{\alpha}
    +12gv_{a}  A\psi^{\alpha}\psi_{\alpha}\nonumber\\
    &&+\left(4g v_{a}^3 +v_{f} \right) F + 12g v_{a}^2v_{f} A +\frac{1}{2}v_{f}^2
    +4g v_{f}v_{a}^3+\mathcal{L}_{CT}\Big]~.
\end{eqnarray}

\noindent
Again, the auxiliary field $F$ can be eliminated using its equation of motion, 
\begin{eqnarray}\label{seq5}
F + 4g A^3 +12gv_{a} A^2 + 12gv_{a}^2 A + 4gv_{a}^3+v_{f}  =0~,
\end{eqnarray}

\noindent
thus allowing us to rewrite Eq.(\ref{seq4}) only as function of the physical fields $A$ and $\psi$,
\begin{eqnarray}\label{seq6}
S&=&\int{d^3x}\Big[\frac{1}{2}A\Box A +\frac{1}{2}\psi^{\alpha}i{\partial_{\alpha}}^{\beta}\psi_{\beta}
+6gv_{a}^2\psi^{\alpha}\psi_{\alpha} 
- 8g^2 A^6 - 48g^2v_{a} A^5 -120g^2v_{a}^2 A^4\nonumber\\ 
&-&160g^2v_{a}^3 A^3 -120g^2v_{a}^4 A^2 
+ 6g A^2\psi^{\alpha}\psi_{\alpha} - 48g^2v_{a}^5 A
+12gv_{a} A\psi^{\alpha}\psi_{\alpha}-8g^2v_{a}^6+\mathcal{L}_{CT} \Big].
\end{eqnarray}

\noindent
Notice the complete cancellation of the vaccum expectation value $v_{f}$. The action above is invariant over SUSY transformations on mass shell. Also, Eq.(\ref{seq6}) is not invariant under the discrete transformation $A(x)\rightarrow -A(x)$, for any value of $v_a\neq 0$. 

The propagators of the model are given by
\begin{eqnarray}\label{seq6a}
\Delta(k)&=&-\frac{i}{k^2+M_{A}^2}~,\nonumber\\
S_{\alpha\beta}(k)&=&\frac{i}{2}\frac{k_{\alpha\beta}-M_{\psi}C_{\alpha\beta}}{k^2+M_{\psi}^2}~,
\end{eqnarray}

\noindent
where $M_A^2=240g^2v_{a}^4$ and $M_{\psi}^2=144g^2v_{a}^4$.

\begin{figure}[ht]
\includegraphics[]{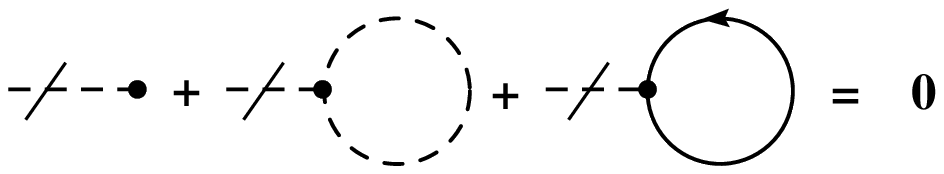}
\caption{One loop tadpole equation. Dashed lines represents the scalar field $A$ propagator, while solid lines represent the fermion field propagator.}\label{f1}
\end{figure}

Now, let us use the tadpole method~\cite{Weinberg:1973ua} to compute loop corrections to the classical potential. The one loop corrections to the tadpole equation are shown in Figure \ref{f1}. The corresponding expressions can be cast as 
\begin{eqnarray}\label{seq8}
\Gamma^{(1)}_{(0+1)l}&=&-48ig^2v_{a}^5
-12gv_{a} M_{\psi}\int\frac{d^3k}{(2\pi)^3}\frac{1}{(k^2+{M_{\psi}}^2)}
-480g^2v_{a}^3 \int\frac{d^3k}{(2\pi)^3} \frac{1}{(k^2+M_A^2)}~.
\end{eqnarray}

\noindent
Integrating over $k$ using regularization by dimensional reduction, we obtain 
\begin{eqnarray}\label{seq9}
\Gamma^{(1)}_{(0+1)l}=\frac{48i}{\pi}\left[(9+10\sqrt{15})g-\pi\right]~g^2~v_{a}^5~.
\end{eqnarray}

\noindent
The effective potential is evaluated from the tadpole equation through the relation
\begin{eqnarray}\label{seq15a}
V_{eff}=i\int_{0}^{A_{cl}}{dv_a}~\Gamma^{(1)}~,
\end{eqnarray}

\noindent
where the integration limits was chosen so that the classical potential vanish at $A_{cl}=0$. 

Integrating Eq.(\ref{seq9}) over $v_a$, we obtain the following effective potential up to one loop order
\begin{eqnarray}\label{seq9b}
V_{eff}&=&i\int_{0}^{A_{cl}}{dv_a}~\Gamma^{(1)}_{(0+1)l}\nonumber\\
&=&\frac{8}{\pi}g^2\left[\pi-(9+10\sqrt{15})g\right]~A_{cl}^6~.
\end{eqnarray}

\noindent
The condition that minimizes the effective potential is obtained by setting $\partial V_{eff}/\partial A_{cl}\Big{|}_{A_{cl}=v_a}=0$, where $v_a$ is the minimum of the effective potential. It is easy to check that $v_a$ possess only the trivial solution $v_a=0$, corroborating the conclusions of~\cite{Burgess:1983nu}. As a next step, we will evaluate the two loop corrections to the effective potential, searching for the dynamical generation of mass. 

The diagrams that contribute to the tadpole equation at two loop order are shown in Figure \ref{f2}, and the mathematical expressions for each diagram are evaluated in appendix \ref{2lcalc}. The effective potential at the two loop approximation is given by
\begin{eqnarray}\label{seq16aa}
V_{eff}=-\frac{d}{6}g^4~A_{cl}^6~\ln\left[e^{(a/g^2+b/g+c-d/3)/d}~\frac{A_{cl}^2}{\mu}\right]+BA_{cl}^6~,
\end{eqnarray}

\noindent
where $B$ is a convenient counter-term, $\mu$ is the mass parameter introduced by the regularization by dimensional reduction, and the constants $a$, $b$, $c$ and $d$ are given by
\begin{eqnarray}\label{seq16aaa}
&&a=-48~,\hspace{2cm} b=(480\sqrt{15}-864)/\pi\approx 317~,\nonumber\\ c&=&[720\sqrt{15}+7182(1+\gamma+1/\epsilon)-14400\ln(12\sqrt{15}g)+36\ln(24g+4g\sqrt{15})]/\pi^2\nonumber\\
&\approx& -(6111+1457\ln{g}-728/\epsilon)~,\nonumber\\ 
&&d=-14364/\pi^2\approx -1457~, 
\end{eqnarray}

\noindent
with $\gamma$ being the Euler's constant and $\epsilon=3-D$ ($D$ is the dimension of the spacetime).

Only one renormalization condition is necessary to ensure the renormalizability of the model. Let us  define the renormalized coupling constant as
\begin{eqnarray}\label{seq17}
\frac{\partial^6 V_{eff}}{\partial A_{cl}^6}\Big|_{A_{cl}=v_a}\equiv
\frac{\partial^6 V_{tree}}{\partial A_{cl}^6}= 8\times6!~g^2~,
\end{eqnarray}

\noindent
resulting in the following expression for the counter-term:
\begin{eqnarray}\label{seq18}
B=\frac{g^2}{60}\left[480+49dg^2\ln\left( e^{a/g^2+b/g+c-d/3}\frac{v^2}{\mu}\right)\right]~.
\end{eqnarray}

\noindent
Therefore, substituting the value of the counter-term $B$ into Eq.(\ref{seq16aa}), the renormalized effective potential can be cast as
\begin{eqnarray}\label{seq18a}
V_{effR}=8g^2A_{cl}^6+\frac{d}{60}g^4A_{cl}^6\left[49-10\ln\frac{A_{cl}^2}{v_a^2}\right]~.
\end{eqnarray}

\noindent
The point of minimum is obtained from the first derivative of $V_{effR}$ with respect to $A_{cl}$, resulting in
\begin{eqnarray}\label{seq18b}
A_{cl}=\pm v_a~\exp\left(\frac{137}{60}+\frac{24}{dg^2}\right)~.
\end{eqnarray}

\noindent
We expect that the effective potential represents a good approximation for values of $A_{cl}\sim v_a$. Therefore we can see that the exponential must be approximately $1$, constraining $g^2$ to be of order of $g^2\sim -1440/(137d)\approx 0.7\times 10^{-2}$. So, we are obtaining a coupling constant $g^2\ll 1$, validating the hypothesis taken in perturbation theory.  

The mass of scalar field $A$ can be defined by
\begin{eqnarray}\label{seq18c}
m_A^2\equiv\frac{\partial^2 V_{effR}}{\partial A_{cl}^2}\Big|_{A_{cl}=v_a} 
=-2dg^4v_a^4\approx 2914g^4v_a^4~>~0~,
\end{eqnarray}

\noindent
where the ratio between the $A$ and $\psi$ masses is ${m_A^2}/{m_{\psi}^2}\approx 1,7$.  
 
In conclusion, we have shown that the SUSY $\Phi^4$ model exhibits dynamical supersymmetry breaking and mass generation at two loop order, in three dimensional spacetime. This was established by using the tadpole method, computing the two loop corrections to the effective potential and showing that its minimum no longer happens at $\langle A\rangle=0$, that is, the quantum corrections makes the point of classical minimum $A_{cl}=0$ a local maximum, dislocating the minimum to a non-trivial value of $A_{cl}$. At the corrected minimum, $A_{cl}=v_a$, the effective potential is non-vanishing, indicating that supersymmetry is broken, and the masses of bosonic and fermionic fields are different, their ratio being ${m_A^2}/{m_{\psi}^2}\approx 1,7$. We also showed that the new minimum, differently from non-supersymmetric models, is compatible with the hypothesis that $g\ll1$, thus validating our perturbative calculations. We can see that the supersymmetry breaking is induced by the dynamical breaking of the discrete symmetry $A(x)\rightarrow -A(x)$, and so no goldstino appears in this model.

\appendix

\section{Two loop calculations}\label{2lcalc}

\begin{figure}[ht]
\includegraphics[]{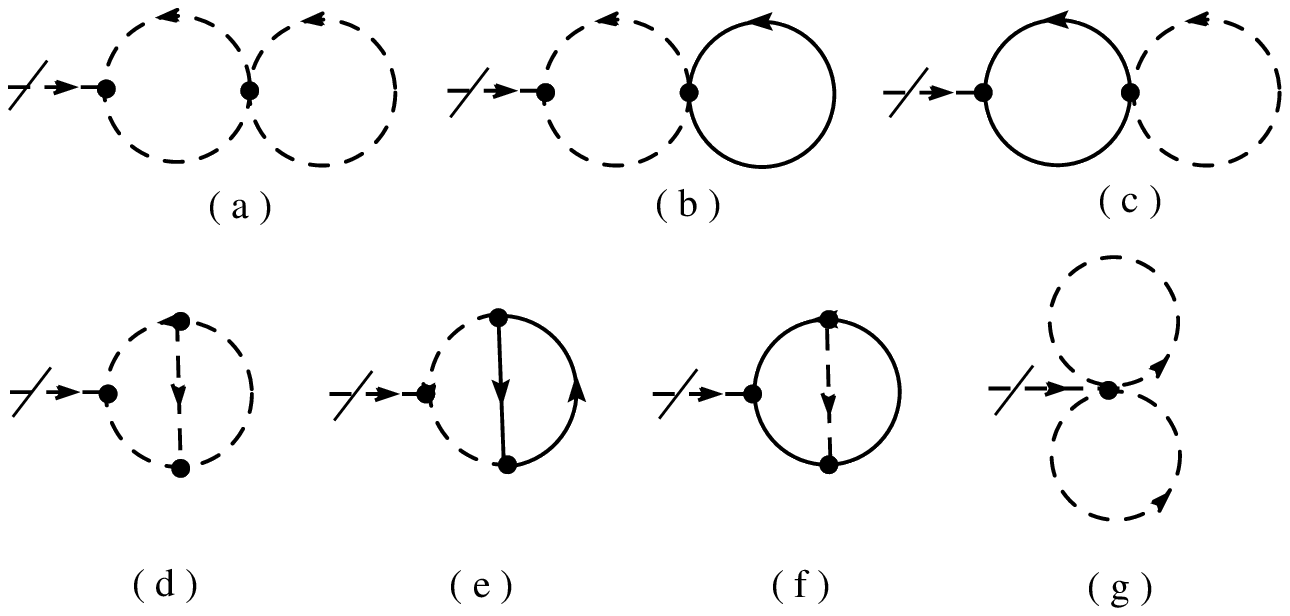}
\caption{Two loop contributions to the tadpole equation.}\label{f2}
\end{figure}

The contributions at two loop to the tadpole equation, Figure \ref{f2}, are given by  
\begin{eqnarray}\label{seq13a}
\Gamma^{(1)}_{2l}(a) & = &-345600 i g^4 v_a^5 \int\frac{d^3k}{(2\pi)^3}\frac{1}{(k^2+M_A^2)^2}
\int\frac{d^3q}{(2\pi)^3}\frac{1}{q^2+M_{A}^2}~,
\end{eqnarray}
\begin{eqnarray}\label{seq13b}
\Gamma^{(1)}_{2l}(b) & = &-2880 i g^3 v_a^3 \int\frac{d^3k}{(2\pi)^3}\frac{1}{(k^2+M_A^2)^2}
\int\frac{d^3q}{(2\pi)^3}\frac{M_{\psi}}{q^2+M_{\psi}^2}~,
\end{eqnarray}
\begin{eqnarray}\label{seq13c}
\Gamma^{(1)}_{2l}(c) & = &72 i g^2 v_a \int\frac{d^3k}{(2\pi)^3}\frac{d^3k}{(2\pi)^3}\Big[
\frac{2M_{\psi}}{(k^2+M_{\psi}^2)^2(q^2+M_A^2)}-
\frac{1}{(k^2+M_{\psi}^2)(q^2+M_A^2)}\Big]~,
\end{eqnarray}
\begin{eqnarray}\label{seq13d}
\Gamma^{(1)}_{2l}(d) & = &-12i(160)^3 g^6 v_a^9 \int\frac{d^3k}{(2\pi)^3}
\int\frac{d^3q}{(2\pi)^3}\frac{1}{(k^2+M_A^2)^2[(k+q)^2+M_A^2](q^2+M_{A}^2)}~,
\end{eqnarray}
\begin{eqnarray}\label{seq13e}
\Gamma^{(1)}_{2l}(e) & = &-23040 i g^4 v_a^5 \int\frac{d^3k}{(2\pi)^3}
\int\frac{d^3q}{(2\pi)^3}\frac{k.q+q^2-M_{\psi}^2}{(k^2+M_A^2)^2[(k+q)^2+M_{\psi}^2](q^2+M_{\psi}^2)}~,
\end{eqnarray}
\begin{eqnarray}\label{seq13g}
\Gamma^{(1)}_{2l}(f) = 0~,~~~~~\Gamma^{(1)}_{2l}(g)  = 960 i g^2 v_a \int\frac{d^3k}{(2\pi)^3}\frac{1}{k^2+M_A^2}
\int\frac{d^3q}{(2\pi)^3}\frac{1}{q^2+M_{A}^2}~,
\end{eqnarray}
\begin{eqnarray}\label{seq13h}
\Gamma^{(1)}_{2l}(h) & = &-230400 i g^4 v_a^5 \int\frac{d^3k}{(2\pi)^3}
\int\frac{d^3q}{(2\pi)^3}\frac{1}{(k^2+M_A^2)^2[(k+q)^2+M_A^2](q^2+M_{A}^2)}~,
\end{eqnarray}
\begin{eqnarray}\label{seq13i}
\Gamma^{(1)}_{2l}(i) & = &72 i g^2 v_a \int\frac{d^3k}{(2\pi)^3}
\int\frac{d^3q}{(2\pi)^3}\frac{k.q-M_{\psi}^2}{(k^2+M_{\psi}^2)[(k+q)^2+M_A^2](q^2+M_{\psi}^2)}~.
\end{eqnarray}

Resolving the integrals in Eqs.(\ref{seq13a}-\ref{seq13i}) with the help of formulae~\cite{Tan:1996kz,Tan:1997ew,Dias:2003pw}, where was adopted regularization by dimensional reduction, and adding the tree, one, and two loop contributions, we obtain the following expression to the tadpole equation:
\begin{eqnarray}\label{seq14}
\Gamma^{(1)}_{(0+1+2)l} & = & i a g^2 v_a^5 +ibg^3v_a^5+icg^4v_a^5+idg^4v_a^5\ln\left(\frac{v_a^2}{\mu}\right)~,
\end{eqnarray}

\noindent
where $a=-48$, $b=(480\sqrt{15}-864)/\pi\approx 317$, $c=[720\sqrt{15}+7182(1+\gamma+1/\epsilon)-14400\ln(12\sqrt{15}g)+36\ln(24g+4g\sqrt{15})]/\pi^2\approx -(6111+1457\ln{g}-728/\epsilon)$ and $d=-14364/\pi^2\approx -1457$, with $\gamma$ being the Euler's constant, $\mu$ is the mass parameter introduced by the regularization, and $\epsilon=3-D$ ($D$ is the dimension of the spacetime).

The effective potential is given by
\begin{eqnarray}\label{seq15}
V_{eff}=i\int_{0}^{A_{cl}}{dv_a}~\Gamma_{(0+1+2)l}^{(1)}~,
\end{eqnarray}

\noindent
from which we obtain
\begin{eqnarray}\label{seq16}
V_{eff}=-\frac{d}{6}g^4~A_{cl}^6~\ln\left[e^{(a/g^2+b/g+c-d/3)/d}~\frac{A_{cl}^2}{\mu}\right]+BA_{cl}^6~.
\end{eqnarray}

\noindent
with $B$ being a counter-term. The limits of integration in Eq.(\ref{seq15}) was chosen for that the classical potential vanish at $A_{cl}=0$.

\vspace{1cm}
{\bf Acknowledgments.} The author would like to thank A. F. Ferrari and A. J. da Silva for the careful reading and useful comments. This work was supported by the Brazilian agency Conselho Nacional de Desenvolvimento Cient\'{\i}fico e Tecnol\'{o}gico (CNPq).

 
\begin{thebibliography}{99}

\bibitem{Witten:1982df}
  E.~Witten,
  Nucl.\ Phys.\  B {\bf 202}, 253 (1982).

\bibitem{Witten:1981nf}
  E.~Witten,
  Nucl.\ Phys.\  B {\bf 188}, 513 (1981).

\bibitem{Burgess:1983nu}
  C.~P.~Burgess,
  Nucl.\ Phys.\  B {\bf 216}, 459 (1983).

\bibitem{Coleman:1973jx}
  S.~R.~Coleman and E.~Weinberg,
  Phys.\ Rev.\  D {\bf 7}, 1888 (1973).

\bibitem{Dias:2003pw}
  A.~G.~Dias, M.~Gomes and A.~J.~da Silva,
  Phys.\ Rev.\  D {\bf 69}, 065011 (2004).

\bibitem{Tan:1996kz}
  P.~N.~Tan, B.~Tekin and Y.~Hosotani,
  Phys.\ Lett.\  B {\bf 388}, 611 (1996).

\bibitem{Tan:1997ew}
  P.~N.~Tan, B.~Tekin and Y.~Hosotani,
  Nucl.\ Phys.\  B {\bf 502}, 483 (1997).
 
\bibitem{Gates:1983nr}
  S.~J.~Gates, M.~T.~Grisaru, M.~Rocek and W.~Siegel,
  Front.\ Phys.\  {\bf 58}, 1 (1983).

\bibitem{Grisaru:1979wc}
  M.~T.~Grisaru, W.~Siegel and M.~Rocek,
  Nucl.\ Phys.\  B {\bf 159}, 429 (1979).

\bibitem{Weinberg:1973ua}
  S.~Weinberg,
  Phys.\ Rev.\  D {\bf 7}, 2887 (1973).


\end{thebibliography}
\end{document}